\documentclass[12pt]{article}
\pdfoutput=1
\usepackage{latexsym}
\usepackage[nosort]{cite}
\usepackage{epsfig,amssymb,euscript,slashed}
\usepackage[normalem]{ulem} 
\usepackage{amsmath}
\usepackage{float}
\usepackage[linktocpage]{hyperref}
\usepackage{ytableau}
\usepackage{bbm}
\usepackage{braket}
\usepackage{xcolor}
\usepackage{fancybox}
\usepackage{geometry}
\if{}
\fi

\def\be{\begin{equation}}
\def\ee{\end{equation}}
\def\bseq{\begin{subequations}}
\def\eseq{\end{subequations}}

\def\bea{\begin{eqnarray}}
\def\eea{\end{eqnarray}}

\def\bseq{\begin{subequations}}
\def\eseq{\end{subequations}}

\numberwithin{equation}{section} 
\usepackage[]{graphicx}

\newcommand {\cH}{{\cal H}}

\newcommand {\cV}{{\cal V}}

\newcommand{\non}{\nonumber}

\def\tr           {\mathop{\rm Tr}}

\def\sqr#1#2{{\vcenter{\vbox{\hrule height.#2pt
 \hbox{\vrule width.#2pt height#1pt \kern#1pt \vrule width.#2pt}\hrule
 height.#2pt}}}}

\def\a{\alpha}
\def\b{\beta}
\def\r{\rho}

\def\g{\gamma}

\def\m{\mu}
\def\n{\nu}

\def\r{\rho}
\def\s{\sigma}

\def\L{\Lambda}

\def\m{\mu}
\def\g{\gamma}
\def\n{\nu}

\newcommand{\ve}{\varepsilon}

\newcommand{\pa}{\partial}
\newcommand{\hf}{\frac12}

\def\slashchar#1{\setbox0=\hbox{$#1$}           
\dimen0=\wd0                                 
\setbox1=\hbox{/} \dimen1=\wd1               
\ifdim\dimen0>\dimen1                        
\rlap{\hbox to \dimen0{\hfil/\hfil}}      
#1                                        
\else                                        
\rlap{\hbox to \dimen1{\hfil$#1$\hfil}}   
/                                         
\fi}

\begin{document}

\begin{titlepage}
\begin{flushright}
\end{flushright}
\vspace{5mm}

\begin{center}
{\Large \bf
On non-linear chiral 4-form theories in $D=10$}
\end{center}

\begin{center}

{\bf
Jessica Hutomo,${}^{a}$
Kurt Lechner,${}^{b, a}$
Dmitri P. Sorokin ${}^{a, b}$
} \\
\vspace{5mm}

\footnotesize{
${}^{a}$
{\it
INFN, Sezione di Padova, Via Marzolo 8, 35131 Padova, Italy
}
 \\~\\
${}^{b}$
{\it
Dipartimento di Fisica e Astronomia “Galileo Galilei” \\
Universit\`a degli Studi di Padova, Via Marzolo 8, 35131 Padova, Italy
}
}
\vspace{5mm}
~\\
\texttt{hutomo@pd.infn.it,
lechner@pd.infn.it,
sorokin@pd.infn.it}\\
\vspace{2mm}
\end{center}

\begin{abstract}
\baselineskip=14pt
\noindent
We consider properties of non-linear theories of a chiral 4-form gauge field $A_4$ in ten space-time dimensions with an emphasis on a subclass of these theories which are invariant under the $D = 10$ conformal symmetry. We show that general results regarding a peculiar structure of duality-invariant abelian gauge theories in four and six space-time dimensions do not extend to non-linear chiral 4-form theories in ten dimensions. This discrepancy arises primarily from the large number 81 of independent invariants constructible from the self-dual part of the five-form field strength $F_5=dA_4$ in $D=10$, in stark contrast to the single independent (fourth-order) invariants which are building blocks of the actions in the lower dimensional cases. In particular, unlike the $D=4$ and $D=6$ cases, where non-linear duality-invariant theories can be viewed as stress-tensor ($T\overline T$-like) deformations of seed theories, the flow equations in $D=10$ generally involve both stress-tensor invariants and additional higher-order structures constructed from $F_5$. In passing, we prove the equivalence of three Lagrangian formulations of non-linear duality-invariant $p$-form theories: the PST, the Ivanov-Nurmagambetov-Zupnik and the ``clone" one.

\end{abstract}
\vspace{5mm}

\end{titlepage}


\newpage
\renewcommand{\thefootnote}{\arabic{footnote}}
\setcounter{footnote}{0}

\tableofcontents{}
\vspace{1cm}
\bigskip\hrule

\allowdisplaybreaks
\section{Introduction}
In this paper we will consider properties of  non-linear theories of a chiral 4-form gauge field in ten space-time dimensions with an emphasis on a subclass of these theories which are invariant under the $D=10$ conformal symmetry. In contrast to the cases of $D=4$ non-linear duality-invariant electrodynamics and a $D=6$ chiral 2-form gauge theory in which there exist unique non-linear conformal deformations of $D=4$ Maxwell's theory and its $D=6$
chiral 2-form counterpart \cite{Bandos:2020jsw,Bandos:2020hgy},
\footnote{These conformal theories are unique in the class of the theories (considered in this paper) that do not contain derivatives of the field strength, a restriction which is natural for avoiding Ostrogradski ghosts. If one allows field-strength derivatives in the Lagrangian, then the class of conformal duality-invariant theories becomes much larger \cite{Kuzenko:2024zra}. The general formalism to describe duality-invariant theories with higher derivatives was given in \cite{Kuzenko:2000uh}.}
in $D=8$ and 10 there is a wide class of duality-invariant conformal theories of $(\frac D 2 -1)$-form gauge fields \cite{Bandos:2020hgy,Avetisyan:2022zza}.

The reason behind this difference is that in
$D=4$ and $D=6$ there exists only one independent duality-invariant Lorentz scalar that one can construct with the components of a (twisted) self-dual\footnote{In space-time dimensions $D=4n+2$ of Lorentz signature there exist real self-dual forms $F_{D/2}$, while in $D=4n$ the Hodge duality relates two different real $F_{D/2}$ forms (or one complex). This leads to the notion of twisted self-duality. It was introduced within a double field formalism for the bosonic sector of the maximal supergravities\cite{Cremmer:1998px}. Here by ``twisted" self-duality we simply mean that a complex $2n$-form in $D=4n$ satisfies the condition $\tilde F_{2n}=\pm iF_{2n}$, where `tilde' denotes the Hodge-dual form and the imaginary unit plays the role of the``twist".} $D/2$-form field strength $F_{D/2}=dA_{(D/2-1)}$ of the gauge field. This invariant is of the fourth order in the components of $F_{D/2}$. In $D=8$ and $D=10$ the situation changes drastically, since the number of functionally independent higher-order invariants of a (twisted) self-dual $F_{D/2}$ becomes very large. There are  41 invariants of this kind  in $D=8$ and 81 invariants in $D=10$ (the counting is based on an algebraic reasoning \cite{Cederwall:2025ywy}).
\if{}
\footnote{The counting is as follows. A four-form $F_4$ in $D=8$ has 70 components, It can be shown that there is no a subgroup of the 28-dimensional Lorentz group $SO(1,7)$ which leaves $F_4$ invariant. Hence, the number of functionally independent Lorentz (pseudo)scalars constructible from $F_4$ is 42=70-28. However, in general these scalars are not invariant under duality rotations of $F_4$ and its Hodge dual $\tilde F_4$. $U(1)$ duality symmetry  requires to construct the invariants as real polynomials of twisted self-dual $F_4+i\tilde F_4$ and its complex conjugate anti-self-dual $F_4-i\tilde F_4$. This further reduces the number to 41 \cite{Cederwall:2025ywy}.}.
\fi
To the best of our knowledge, by now  there is no complete classification of these invariants in the literature due to the complexity of this group-theoretical problem, especially in $D=10$.\footnote{See \cite{Cederwall:2025ywy} for details on the analysis of invariants of tensor fields in various dimensions.} Only very limited results on the structure of these invariants at the 4th \cite{Buratti:2019cbm,Buratti:2019guq,Avetisyan:2022zza} and 8th-order \cite{Paulos:2008tn,Melo:2020amq,Liu:2022bfg} were obtained (in \cite{Paulos:2008tn,Melo:2020amq} an explicit form of five 8th-order invariants of the self-dual 5-form were individualized within eight-derivative $\alpha'{}^3$-corrections to the effective action of type IIB supergravity).

Because of the huge number of independent invariants in $D=8$ and $10$, also the structure of generic duality-invariant non-linear theories of $(D/2-1)$-form gauge fields becomes much more complicated in comparison to the $D=4$ and 6 cases, in which the interacting part of the actions is a generic functional of a single quartic duality-invariant Lorentz scalar $(F_{D/2})^4$ \cite{Kuzenko:2000uh,Ivanov:2002ab,Ivanov:2003uj,Kuzenko:2013gr,Ivanov:2014nya,Buratti:2019cbm,Buratti:2019guq,Kuzenko:2019nlm,Avetisyan:2021heg,Avetisyan:2022zza}.\footnote{This is related to the fact that the form of non-linear duality-invariant Lagrangians in $4D$ and $6D$ theories is governed by solutions of a Courant-Hilbert equation \cite{C-H}, which are implicit functions of a single variable (see e.g. \cite{Gibbons:1995cv,Perry:1996mk,Gaillard:1997rt,Hatsuda:1999ys,Russo:2024llm,Russo:2025wph,Russo:2025fuc,Babaei-Aghbolagh:2025cni}).} As was recently shown in \cite{Ferko:2023wyi,Ferko:2024zth} (see \cite{Kuzenko:2025jgk} for a review), all duality-invariant non-linear electrodynamics in $D=4$ and all non-linear chiral 2-form theories in $D=6$ (except for the non-linear conformal electrodynamics of the Bia\l{}ynicki-Birula type \cite{Bialynicki-Birula:1984daz,Bialynicki-Birula:1992rcm,Gibbons:2000ck,Townsend:2019ils}) can be regarded as stress-tensor deformations of the free (Maxwell or chiral 2-form) theory, or, more generally, of  non-linear duality-invariant seed theories. The  non-linear Lagrangian densities $\mathcal L(F)$ of the deformed theories obey the following flow equation
\begin{align}\label{lagrangian_flow_general}
    \frac{\partial \mathcal{L}^{(\lambda)}}{\partial \lambda} = f \left( T_{\mu \nu}^{(\lambda)} , \lambda \right) \, ,
\end{align}
with the initial condition $\mathcal L|_{\lambda=0}=\mathcal L^{seed}$,
where $\lambda$ is a parameter of the deformation and $f$ is a duality invariant function of the energy-momentum tensor  $T^{(\lambda)}_{\mu\nu}$ of the deformed theory. In this paper we will show (on examples) that this is not so for the generic non-linear theory of the chiral 4-form in $D=10$.

We will start with the consideration of the general structure of the non-linear $D=10$ chiral 4-form theory and then concentrate on the study of examples in which non-linearities in the action are determined by the unique fourth-order invariant \cite{Buratti:2019guq,Avetisyan:2022zza}
which one can construct with the components of a self-dual 5-form $\Lambda_5$ which on the mass shell coincides with the self-dual part $F^+_5=\tilde F^+_5$ of the physical field strength\,, namely
\be\label{I4}
I_4=M_{\mu}{}^{\nu}M_{\nu}{}^{\mu}=\tr M^2:=M^2, \qquad M_{\mu}{}^{\nu} :=\Lambda_{\mu\rho_1\rho_2\rho_3\rho_4}\Lambda^{\nu\rho_1\rho_2\rho_3\rho_4}\,,
\ee
where the Greek letters stand for the indices of $10D$ space-time of almost plus signature $(-,+,\ldots,+)$. We will see that already this simplest subclass of non-linear chiral 4-form theories is much more complicated than its 6-dimensional counterpart.

For the description of the chiral 4-form theory we will use the formulation developed in \cite{Ferko:2024zth} as a generalization to arbitrary space-time dimensions of the construction of Ivanov, Nurmagambetov and Zupnik (INZ) \cite{Ivanov:2014nya} who unified  the covariant PST formulation \cite{Pasti:1995ii,Pasti:1995tn} with the auxiliary tensor field construction of \cite{Ivanov:2002ab,Ivanov:2003uj} to describe non-linear duality-symmetric electrodynamics in $D=4$. We find this INZ-type formulation to be the most suitable for the purposes of this paper.\footnote{ For a review of other Lagrangian formulations to the description of duality-invariant theories see e.g. \cite{Bandos:2020hgy,Evnin:2022kqn,Sorokin:2025ezs}.} In Section \ref{INZvsPST} we will refine the proof of the relation between the INZ-type and the PST Lagrangian formulation \cite{Buratti:2019guq,Bandos:2020hgy} of the interacting theories of chiral $2n$-forms in $D=4n+2$ given in \cite{Ferko:2024zth}. In Section \ref{INZvsclone}, using a simple field redefinition, we prove the full equivalence of the INZ construction and a more recent formulation of non-linear duality-invariant $p$-form theories dubbed ``clone" \cite{Evnin:2022kqn}. The latter was introduced { for free theories} by Mkrtchyan \cite{Mkrtchyan:2019opf} and generalized to non-linear theories in \cite{Avetisyan:2021heg}, \cite{Avetisyan:2022zza}.\footnote{ An attempt to relate (at the Lagrangian level) the PST formulation to yet another (Sen's) approach \cite{Sen:2015nph,Sen:2019qit} to the description of chiral $p$-form theories  was undertaken in \cite{Vanichchapongjaroen:2024tkj}.}

\section{INZ-type approach to the description of interacting chiral 4-forms in $D=10$}

In this section we will apply the formulation of \cite{Ivanov:2014nya,Ferko:2024zth}  to the construction of an action for the non-linear chiral 4-form theory in a curved $D=10$ space-time. The construction involves the field strength $F_5=dA_4$ of the physical 4-form gauge field $A_4(x)$ (which {\it apriori} is not subject to any self-duality condition) and an auxiliary self-dual 5-form field $\Lambda_5(x)=\tilde \Lambda_5(x)$, i.e.
\be\label{L=tildeL}
\Lambda^{\mu_1\ldots\mu_5}=\frac{1}{5!\sqrt{-g}}\varepsilon^{\mu_1\ldots\mu_5\nu_1\ldots\nu_5}\Lambda_{\nu_1\ldots\nu_5}\,
\ee
with $\varepsilon^{\mu_1\ldots\mu_5\nu_1\ldots\nu_5}$ being the $10D$ Levi-Civita density with $\ve^{01\dots 9} = -1$.

The construction also involves an auxiliary scalar $a(x)$, which entails the unit time-like vector
\be\label{v}
v_\mu=\frac{\partial_\mu a}{\sqrt{-(\partial a)^2}}, \qquad v_\mu v^\mu=-1\,.
\ee
To build the action we will use the components of $F_5$ and $\Lambda_5$ contracted with $v_\mu$, which we denote as follows
\be\label{EB}
E_{\mu_1\mu_2\mu_3\mu_4}=F_{\mu_1\mu_2\mu_3\mu_4\rho}v^{\rho}, \qquad B_{\mu_1\mu_2\mu_3\mu_4}=\tilde F_{\mu_1\mu_2\mu_3\mu_4\rho}v^{\rho}\,,
\ee
\be\label{lambda}
\lambda_{\mu_1\mu_2\mu_3\mu_4}=\Lambda_{\mu_1\mu_2\mu_3\mu_4\rho}v^{\rho}=\tilde \Lambda_{\mu_1\mu_2\mu_3\mu_4\rho}v^{\rho}\,.
\ee
For future use we note that the following identities hold
\be
F_{\mu_1\mu_2\mu_3\mu_4\mu_5} = -5 E_{[\m_1 \dots \m_4} v_{\m_5]}-\frac{1}{4!}\sqrt{-g} \ve_{\m_1 \dots \m_5 \n_1 \dots \n_5}B^{\n_1 \dots \n_4} v^{\n_5}~,
\label{F=E+B}
\ee
and
\be
\Lambda_{\mu_1\mu_2\mu_3\mu_4\mu_5} = -5 \lambda_{[\mu_1 \dots \mu_4} v_{\mu_5]}-\frac{1}{4!}\sqrt{-g} \ve_{\mu_1 \dots \mu_5 \nu_1 \dots \nu_5}\lambda^{\nu_1 \dots \nu_4} v^{\nu_5}~.
\label{Lambda-dec}
\ee

The action has the following form
\be
\mathcal S[A, a, \L] 
= \frac{1}{4!} \int d^{10}x \sqrt{-g} \,\Big( \hf E \cdot B- \hf B \cdot B + (B+\lambda) \cdot (B+ \lambda) -\mathcal V(\Lambda)\Big)~,
\label{INZ}
\ee
where dot denotes the scalar product of the tensors and $\mathcal V(\Lambda)$ is an arbitrary Lorentz-invariant function of $\Lambda_5$, which determines the form of interactions in the theory. Note that since $\Lambda_5$ is self-dual, $\Lambda \cdot \Lambda \equiv 0$ and hence $\mathcal V(\Lambda)$ can depend only on invariants of $\Lambda_5$ starting from the order four and higher.

When $\mathcal V(\Lambda)=0$, the equation of motion of $\Lambda_5$ obtained by the variation of the action \eqref{INZ} reduces to
\be\label{l=-B}
\lambda=-B.
\ee
Upon the substitution of this solution into the action \eqref{INZ} (with $\mathcal V(\Lambda)=0$), the latter reduces to
\be\label{PST}
\mathcal S_{\rm free}^{\rm PST} [A, a]=\frac{1}{2\cdot 4!} \int d^{10}x \sqrt{-g} \Big(  E \cdot B-  B \cdot B\Big)\,,
\ee
which is the action for a free chiral 4-form gauge field in the PST formulation \cite{Pasti:1996vs,Dall'Agata:1997ju,Dall'Agata:1998va}. { A local symmetry of the PST action allows one to impose the gauge fixing condition $\partial_{\mu}a(x)=\delta_\mu^0$. In this case the action reduces to the non-manifestly space-time invariant Hamiltonian-type action of Henneaux and Teitelboim \cite{Henneaux:1988gg}.}
The equations of motion of $A_4$ derived from this action reduce to the Hodge self-duality condition on the physical field strength (see \cite{Sorokin:2025ezs} for a recent review)
\be\label{F=F*}
F_5=\tilde F_5\,.
\ee

\subsection{General structure of chiral 4-form interactions}
As we have mentioned, the interaction term $\mathcal V(\Lambda)$ is a space-time invariant function of the components of $\Lambda_5$.
In total there are 81 functionally independent invariants which one can construct from different powers of the components of the self-dual 5-form \cite{Cederwall:2025ywy}. The counting is as follows. The $10D$ self-dual 5-form $\Lambda_5$ has 126 independent components. It can be shown that the $10D$ Lorentz group $SO(1,9)$, whose dimension is 45, acts non-trivially on $\Lambda_5$, i.e. there is no  subgroup of $SO(1,9)$ which leaves $\Lambda_5$ invariant. Applying the Lorentz group transformations one reduces the number of independent components of $\Lambda_5$ to $126-45=81$. Thus, the maximal number of functionally independent Lorentz invariants which one can construct from the components of $\Lambda_5$ is 81. Finding an explicit form of all these invariants turns out to be an open and highly non-trivial group-theoretical problem. We will address this issue in detail in a separate publication \cite{Cederwall:2025ywy}, while listing here only a few of the simplest results relevant for further discussion.

First of all, no invariants can be constructed from the odd powers of $\Lambda_5$, and the invariant $\Lambda\cdot\Lambda$ is identically zero.
Next, as was found in \cite{Buratti:2019guq}, \cite{Avetisyan:2022zza}, there is only one independent fourth-order invariant \eqref{I4}
\be\label{I4L}
I_4=M_{\mu}{}^{\nu}M_{\nu}{}^{\mu}=\tr M^2, \qquad M_{\mu}{}^{\nu} :=\Lambda_{\mu\rho_1\rho_2\rho_3\rho_4}\Lambda^{\nu\rho_1\rho_2\rho_3\rho_4}\,.
\ee
The matrix $M_{\mu\nu}$ is symmetric and traceless. It is in the {\bf 54} irreducible representation of $SO(1,9)$. Hence it has 9 independent eigenvalues which correspond to the following 9 independent invariants
\be\label{trM^2n}
I_{2n}=\tr M^n, \qquad n=2,3,\ldots, 10\,.
\ee
One may wonder whether in addition to
\be\label{I61}
I^{(1)}_6=\tr M^3
\ee
there are other independent invariants at the 6th order in $\Lambda$. The Lie program \cite{LiE:1992} tells us that there is only one more independent invariant in the symmetric tensor product of six {\bf 126} irreps of $SO(1,9)$. We have found that the second 6th-order invariant can be chosen in the following form
\be\label{I62}
I^{(2)}_6=N^{^{(1050)}}_{[\rho_1\rho_2\rho_3,\rho_4\rho_5]\kappa}\Big(N_{_{(1050)}}^{[\rho_1 \rho_2 \rho_3,}{}_{\alpha_1 \alpha_2 ]\alpha_3}{}N_{_{(1050)}}^{[\rho_4\rho_5\kappa,\alpha_1 \alpha_3 ]\alpha_2}\Big),
\ee
where
\be\label{N}
N^{^{(1050)}}_{[\rho_1\rho_2\rho_3,\alpha_1 \alpha_2 ]\alpha_3}=\Lambda^{\mu\nu}{}_{[\rho_1\rho_2\rho_3}\Lambda_{\alpha_1 \alpha_2 ]\alpha_3\mu\nu}
\ee
is the composite tensor in which the indices $[\rho_1\rho_2\rho_3,\alpha_1 \alpha_2 ]$ are anti-symmetrized. It takes values in the {\bf 1050} irreducible representation of $SO(1,9)$, i.e. it is traceless and Hodge self-dual with respect to its five antisymmetrized indices. Its Young tableau is
$$
\resizebox{0,5cm}{!}{\tiny
\begin{ytableau}
~&~\\
~\\
~\\
~\\
~
\end{ytableau}
}
$$
$N^{^{(1050)}}$ is part of the decomposition of the composite tensor
\be\label{N2}
N_{\mu_1\mu_2\mu_3,\nu_1\nu_2\nu_3}=\Lambda_{\mu_1\mu_2\mu_3\rho\sigma}\Lambda_{\nu_1\nu_2\nu_3}{}^{\rho\sigma}
\ee
into irreducible representations of $SO(1,9)$ \footnote{Note that the following identity holds
$$
N_{\mu_1\mu_2\alpha,}{}^{\nu_1\nu_2\alpha}=\frac 12\delta^{[\nu_1}_{[\mu_1}M_{\mu_2]}{}^{\nu_2]}\,
$$ because of the self-duality of $\Lambda_5$.}
\be\label{Nsplit}
N_{\mu_1\mu_2\mu_3,\nu_1\nu_2\nu_3}= 5N^{^{(1050)}}_{[\mu_1\mu_2\mu_3,{\color{red}[}\nu_1\nu_2]\nu_3{\color{red}]}}+N^{^{(4125)}}_{\mu_1\mu_2\mu_3,\nu_1\nu_2\nu_3}+\frac{9}{28}\delta^{[\alpha_1}_{[\mu_1}\delta^{\alpha_2}_{\mu_2}M^{^{(54)}}_{\mu_3]}{}^{\alpha_3]}g_{\alpha_1\nu_1}
g_{\alpha_2\nu_2}g_{\alpha_3\nu_3},
\ee
where the anti-symmetrization of the three indices $\nu_i$ with the ``red" brackets is performed after the anti-symmetrization of the five indices $[\mu_1\mu_2\mu_3,\nu_1\nu_2]$ in the ``black" brackets. The tensor $N^{^{(4125)}}_{\mu_1\mu_2\mu_3,\nu_1\nu_2\nu_3}$ is in the {\bf 4125} irrep of $SO(1,9)$, i.e. it is totally traceless, its two groups of the three indices $\mu_i$ and $\nu_i$ are separately antisymmetrized, the tensor is symmetric under the exchange of the groups of the antisymmetrized indices $\mu_i$ and $\nu_i$, while the antisymmetrization of any four of its indices is zero. The Young tableau of this irrep is
$$
\resizebox{0,5cm}{!}
{\tiny
\begin{ytableau}
~&~\\
~&~\\
~&~
\end{ytableau}
}
$$
 At the 8th order in $\Lambda_5$, according to the Lie program \cite{LiE:1992}, there are six independent invariants, among which one can choose
\be\label{I81}
I_8^{(1)}=\tr M^4\,, \qquad I^{(2)}_{8}=N^{^{(4125)}}_{\mu_1\mu_2\mu_3,}{}^{\nu_1\nu_2\nu_3}M_{\nu_1}{}^{\mu_1}M_{\nu_2}{}^{\mu_2}M_{\nu_3}{}^{\mu_3.}\,.
\ee
These two invariants will appear in our examples of interacting chiral 4-form theories. Obviously, we do not count the invariant of the form $I_8=(I_4)^2$ as independent.

Previously, five 8th-order invariants appeared as $(\alpha')^3$-corrections to the effective action of type IIB supergravity \cite{Paulos:2008tn,Melo:2020amq}. The building blocks of a basis of these invariants are four copies of the tensor \eqref{N} to which the derivative term $\nabla_{\alpha_3}\Lambda_{\rho_1\rho_2\rho_3\alpha_1 \alpha_2 }$ is added (see \cite{Cederwall:2025ywy} for details on the structure of 8th-order invariants).

Let us now pass to the consideration of the properties of the theory described by the action  \eqref{INZ}.

\subsection{Symmetries and equations of motion}
In addition to the standard gauge symmetry $A'_4=A_4+d\alpha_3(x)$, the general action \eqref{INZ} is invariant under the following local transformations
\be\label{PST1}
\delta A_{\mu_1\mu_2\mu_3 \mu_4}=v_{[\mu_1}\Phi_{\mu_2\mu_3\mu_4]}(x)\,,\qquad \delta a(x)=0, \qquad \delta\Lambda_{\mu(5)}=0
\ee
and
\begin{align} \label{PSTGT3}
    \delta A_{\mu (4 ) } = -\frac{\varphi(x)}{\sqrt{ - \partial a \partial a}} \left[ E-B + 2 \left( B + \lambda \right) \right]_{\mu(4)} \, , \qquad \delta a(x) = \varphi(x) \, , \qquad
    \delta \Lambda_{\mu(5)} =0\, ,
\end{align}
where $\Phi_{\mu(3)}(x)$ and $\varphi(x)$ are the local transformation parameters and $\mu(i)$ stands for totally anti-symmetrized indices $\mu(i)=[\mu_1\cdots\mu_i]$.

The first local symmetry \eqref{PST1} ensures that the equation of motion of $A_4$ obtained from \eqref{INZ} reduces to a duality-like relation
\be\label{INZdi}
E=-B-2\lambda,\quad {\rm or\,\,equivalently }\quad E-B=-2(B+\lambda)\,,\quad{\rm or}\quad \lambda=-\frac 12 (E+B)\,,
\ee
while the second local symmetry ensures that $a(x)$ is a Stueckelberg-like auxiliary field whose equation of motion is identically satisfied once the relation \eqref{INZdi} holds. The role of this field is to make the chiral $p$-form Lagrangian manifestly invariant under space-time (Lorentz or general coordinate) transformations.

As we have already mentioned, in the free theory case ($\mathcal V(\Lambda)=0$) the $\Lambda_5$ field equation is \eqref{l=-B} and hence \eqref{INZdi} reduces to $E=B$ which, due to the identity \eqref{F=E+B}, is equivalent to the Hodge self-duality of $F_5$, eq. \eqref{F=F*}.
In the interacting theory the equation of motion of $\Lambda_5$ is\footnote{One might be interested in solving the non-dynamical equation \eqref{Leq} for $\Lambda$ as a function of $B$ (at least perturbatively) and substitute the result back into the action \eqref{INZ} thus getting rid of the auxiliary field $\Lambda$ in the action. As we will show in Section \ref{INZvsPST}, in this way one gets the PST Lagrangian formulation of the non-linear theories of chiral p-forms.}
\be\label{Leq}
(B+\lambda)^{[\mu_1 \dots \mu_4} v^{\mu_5]} - \frac{1}{5!\sqrt{-g}} \ve^{\mu_1 \dots \mu_5 \nu_1 \dots \nu_5} (B_{\nu_1 \dots \nu_4} + \lambda_{\nu_1 \dots \nu_4}) v_{\nu_5} = \frac{\partial \mathcal V(\Lambda)}{\partial\Lambda_{\mu_1 \dots \mu_5}}~.~~~~~~~
\ee
Note that both the left- and the right-hand side of the above equation are anti-self-dual, because they are obtained by varying the action with respect to the self-dual $\Lambda_5$ for which the following identity holds
\bea
\frac{\pa \Lambda_{\n_1 \dots \n_5}}{\pa \Lambda_{\m_1 \dots \m_5}} = \hf \delta_{[\n_1}^{\m_1} \dots \delta_{\n_5]}^{\m_5}- \frac{1}{2\cdot 5!} \ve^{\m_1 \dots \m_5}{}_{\n_1 \dots \n_5}~.
\eea
It then follows from the identities \eqref{F=E+B}, \eqref{Lambda-dec} and the second expression in  \eqref{INZdi} that
\be\label{F^-=V/L}
F^-_{\mu_1\ldots\mu_5}=\frac 12 (F_{\mu_1\ldots\mu_5}-\tilde F_{\mu_1\ldots\mu_5})= 5\frac{\partial \mathcal V(\Lambda)}{\partial\Lambda^{\mu_1 \dots \mu_5}}~.
\ee
On the other hand, from the third expression in \eqref{INZdi} and the identities \eqref{F=E+B} and \eqref{Lambda-dec} it follows that
\be\label{L=F+}
\Lambda_5=-\frac 12(F_5+\tilde F_5)=-F^+_5\,.
\ee
Substituting this relation into \eqref{F^-=V/L} we get the non-linear self-duality condition between the anti-self-dual and self-dual part of $F_5$
\be\label{F^-=VF+}
F^-_{\mu_1\ldots\mu_5}=-5 \frac{\partial \mathcal V(F^+)}{\partial F^+{}^{\mu_1 \dots \mu_5}}~.
\ee
Note that this on-shell equation is a Lorentz-covariant duality relation between the components of the physical field $F_5=dA_4$ only. In this form the generic non-linear duality invariant condition for a chiral $2n$-form in $D=4n+2$ was first obtained  in the ``clone" formulation \cite{Avetisyan:2022zza} using an auxiliary tensor field which differs from that in the INZ formulation by a shift and rescaling (see Section \ref{INZvsclone} for the proof of the equivalence of the INZ and the ``clone" formulation).

Contracting eq. \eqref{F^-=VF+} with $F^{+\mu(5)}$ we get
\be\label{F-F+}
F^-\cdot F^{+}=\frac 12 F\cdot F=-5F^{+}\cdot \frac {\partial \mathcal V(F^+)}{\partial F^+}\,.
\ee
So, in any interacting chiral 4-form theory, on the mass shell the square of the field strength $F_5$ is proportional to the derivative of $\mathcal V(F^+)$ along $F^+$.

\subsection{Energy-momentum tensor}
The energy-momentum tensor of the theory described by the action \eqref{INZ} is
\bea
T_{\mu\nu}&=&-\frac{2}{\sqrt{-g}}\frac{\delta \mathcal S}{\delta g^{\mu\nu}} \nonumber\\
&=& T_{\m \n}^{\rm PST, free}- \frac{1}{4!} \bigg\{
8 \big(B_{ \r(3) (\mu}+ \lambda_{\r(3) (\mu} \big)\big(B_{\nu)}{}^{\r(3)}- 2\lambda_{\nu)}{}^{ \r(3)} \big) -3 g_{\m \n} \big( B_{\r(4)} + \lambda_{\r(4)}\big) \lambda^{\r(4)} \non\\
&&+ (g_{\m \n} + 2 v_{\m}v_{\n}) \big(B_{\r(4)}+ \lambda_{\r(4)}\big)\big(B^{\r(4)}+ \lambda^{\r(4)}\big) + 6 v_{(\m}\Lambda_{\n) \r(4)} \big(B^{\r(4)} + \lambda^{\r(4)}\big)
\bigg\}\nonumber\\
&&- {\frac 1{4!}}g_{\m \n}\Big(\cV- \hf \Lambda_{\rho(5)} \, \frac{\pa \cV}{\pa \Lambda_{\rho(5)}} \Big)
\label{emt-inz-free}
\eea
where
\be
T_{\m \n}^{\rm PST, free} = -\frac{1}{6}p_{(\m} v_{\n)} + \frac{1}{2\cdot 4!} \Big( ( g_{\m \n} +2v_{\m} v_{\n})B_{\r(4)}B^{\r(4)} - 8 B_{(\m}{}^{\r(3)} B_{\n) \r(3)}  \Big)
\label{emt-pst-free}
\ee
is the energy-momentum of the free theory in the PST formulation
and
\bea
p_{\m} &=& -\frac{1}{4 \cdot 4!} \sqrt{-g} \ve_{\m \r_1 \cdots \r_4 \n_1 \cdots \n_5} B^{\rho_1\ldots\rho_4} B^{\n_1 \dots \n_4} v^{\n_5}~, \qquad p_{\mu}v^{\mu} = 0~.
\eea
The indices in the round brackets are symmetrized, e.g. $p_{(\m} v_{\n)}=\frac 12 (p_{\m} v_{\n}+p_{\n} v_{\m})$\,.

To derive the energy-momentum tensor we used the fact that $\Lambda_5$ is Hodge self-dual and hence its variation with respect to $g^{\mu\nu}$ is non-zero, namely
\bea
\delta_{g} \Lambda^{\mu(5)} &=&  \Big( \frac{1}{4} g_{\rho \sigma} \Lambda^{\mu(5)} - \frac{1}{2 \cdot 4!\sqrt{-g}} \ve^{\mu(5)\nu(4)}{}_{(\rho} \Lambda_{\sigma)\nu(4)} \Big)\delta g^{\rho \sigma} \non\\
&=& \Big(-\frac{1}{4} g_{\rho \sigma} \Lambda^{\mu(5)} + \frac{5}{2} \delta_{\rho}{}^{[\mu_1} \Lambda^{\mu_2 \dots \mu_5]}{}_{\sigma} \Big) \delta g^{\rho \sigma}~.
\label{var-lambda}
\eea

The form of the energy-momentum tensor drastically simplifies if we express $B$ in terms of $\lambda$ with the use of eq. \eqref{Leq} projected along $v^{\mu_5}$, which gives
\be\label{B=lvl}
B_{\mu(4)}=-\lambda_{\mu(4)}-5\frac{\partial \mathcal V(\Lambda)}{\partial\Lambda^{\mu(4) \mu_5}}v^{\mu_5}\,.
\ee
Remarkably, the resulting form of the energy-momentum tensor is manifestly space-time invariant without the use of the auxiliary vector $v_\mu(x)$
\be
T_{\mu \nu}(\Lambda) = \frac{1}{2 \cdot 4!} \Big( \Lambda_{\mu}{}^{\rho(4)} \Lambda_{\nu \rho(4)} - {25} \frac{\pa \cV}{ \pa \Lambda^{\mu}{}_{\rho(4)}} \frac{\pa \cV}{\pa \Lambda^{\nu \rho(4)}}\Big)- {\frac1{4!}} g_{\m \n}\Big(\cV- \hf \Lambda_{\rho(5)} \, \frac{\pa \cV}{\pa \Lambda_{\rho(5)}} \Big)~.
\label{emt-INZ-int}
\ee
Note that the energy-momentum tensor of the free chiral 4-form theory ($\mathcal V(\Lambda)=0$) is proportional to the matrix $M_{\mu\nu}$  and hence the quartic invariant \eqref{I4L} is proportional to $T_{\mu\nu}T^{\mu\nu}|_{free}$.

If in \eqref{emt-INZ-int} we now use the on-shell relations \eqref{F^-=V/L} and \eqref{L=F+} we find that
\bea
T_{\mu \nu}(F^+) &=& \frac{1}{2 \cdot 4!} \Big( F^+_{\mu}{}^{\rho(4)} F^+_{\nu \rho(4)} - 25 \frac{\pa \cV}{ \pa F^{+\mu}{}_{\rho(4)}} \frac{\pa \cV}{\pa F^{+\nu \rho(4)}}\Big)- {\frac1{4!}}g_{\m \n}\Big(\cV- \hf F^+_{\rho(5)} \, \frac{\pa \cV}{\pa F^+_{\rho(5)}} \Big)~.\nonumber\\
&&
\eea
In this form the energy-momentum tensor depends only on the self-dual components $F^+_5$ of the physical gauge field. If we take into account the equation of motion \eqref{F^-=VF+}, the energy-momentum tensor can be further rewritten in the following form
\be\label{Tftf}
T_{\mu \nu}=\frac{1}{2 \cdot 4!} F_{(\mu}{}^{\rho(4)} \tilde F_{\nu) \rho(4)}- {\frac1{4!}}g_{\m \n}\Big(\cV- \hf F^+_{\rho(5)} \, \frac{\pa \cV}{\pa F^+_{\rho(5)}} \Big)\,.
\ee
Note that the  terms in the first brackets of \eqref{emt-INZ-int}-\eqref{Tftf} are identically traceless. Therefore, the whole energy momentum tensor is traceless if and only if
\be\label{Tmm=0}
T_{\mu}{}^{\mu}={\frac5{12}}\Big(\cV- \hf \Lambda \cdot \, \frac{\pa \cV(\Lambda)}{\pa \Lambda} \Big)=0.
\ee
This homogeneity condition on the interaction function $\cV(\Lambda)$ singles out a very wide subclass of  non-linear chiral 4-form theories which are conformal. For instance, one may take
\be\label{Vconf}
\mathcal V^{\rm conf}=b\sqrt{I_4}+ \Big(c_1 I^{(1)}_6+c_2I^{(2)}_6\Big)^{\frac 13} +d \Big(I_8\Big)^{\frac 14}+e\, \frac {I_6}{I_4}+ \kappa\, \frac {I_8}{I_6}+\ldots\,,
\ee
where the invariants $I_n$ are given in \eqref{I4L}-\eqref{I62} and \eqref{I81}, and ${b,c,d,e,\kappa,\ldots}$ are dimensionless parameters.

The first term in \eqref{Vconf} is a ten-dimensional counterpart of the square-root interaction functions in the Lagrangians of the unique $4D$ and $6D$ ModMax theories \cite{Bandos:2020jsw,Bandos:2020hgy} constructed in formulations using auxiliary fields \cite{Kuzenko:2021cvx}, \cite{Avetisyan:2022zza,Ferko:2024zth}.  We will consider the conformal 10D model with $\mathcal V^{conf}={  b} \sqrt{I_4}$ in more detail in the next Section.

\section{Simplest non-linear chiral 4-form Lagrangians and stress tensor deformations}
Let us now restrict our consideration to the case in which the interaction term $\mathcal V(\Lambda)$ is a function of the fourth-order invariant \eqref{I4L}
\be\label{VI}
\mathcal V(\Lambda)=\mathcal V(I_4)\,.
\ee
In $D=6$ the function $\mathcal V(I_4)$ defines the most general class of interacting chiral 2-form theories \cite{Ferko:2024zth}, since the quartic invariant constructed with components of the self-dual 3-form $\Lambda_3$ is the unique independent invariant. Therefore, it is interesting to compare the properties of non-linear $10D$ chiral 4-form theories in  this specific class with their 6-dimensional counterparts.

In this case, equation \eqref{F^-=V/L} reduces to
\be\label{F^-=V/L4}
F^-_{\mu_1\ldots\mu_5}=20 M_{[\mu_1}{}^\rho\Lambda_{\mu_2 \mu_3\mu_4 \mu_5]\rho}\mathcal V_{I}~, \qquad \mathcal V_{I}:=\frac {\partial \mathcal V}{\partial I_4}\,
\ee
and the energy-momentum tensor \eqref{emt-INZ-int} takes the form
\bea
T_{\mu \nu}(\Lambda) & =& \frac{1}{2 \cdot 4!} M_{\mu\nu} \Big(1-{24}(\mathcal V_I)^2\,\tr M^2\Big)- \frac 1{4!}g_{\m \n}\Big(\cV- 2I_4 \mathcal V_I \Big)\nonumber\\
&&+\frac {16(\mathcal V_I)^2}{2\cdot 4!} \Big( 2(M^3)_{\mu \nu}-\frac{1}{2} \tr M^3 g_{\mu \nu}
-{12} N_{\mu\alpha_1\alpha_2,\, \nu}{}^{\beta_1\beta_2}M_{\beta_1}{}^{\alpha_1}M_{\beta_2}{}^{\alpha_2} \Big)\,,
\label{emt-INZ-int4}
\eea
where
$N_{\mu\alpha_1\alpha_2,}{}^{\nu\beta_1\beta_2}$ was defined in \eqref{N2}. Using its decomposition \eqref{Nsplit}
and the identity
$$
9\delta_{[\mu}^{[\nu}\delta_{\alpha_1}^{\beta_1}M_{\alpha_2]}{}^{\beta_2]}M_{\beta_1}{}^{\alpha_1}M_{\beta_2}{}^{\alpha_2} =-\Big(\delta_{\mu}^{\nu}\tr M^3-3(M^3)_{\mu}{}^{\nu}+\frac 32 M_{\mu}{}^{\nu}\,\tr M^2\Big)\,,
$$
we can rewrite the energy-momentum tensor \eqref{emt-INZ-int4} as follows
\bea
T_{\mu \nu}(\Lambda) & =& \frac{1}{2 \cdot 4!} M_{\mu\nu} \Big(1-{\frac {96}{7}}(\mathcal V_I)^2\,\tr M^2\Big)- \frac 1{4!} g_{\m \n}\Big(\cV- 2I_4 \mathcal V_I \Big)\nonumber\\
&&+\frac {(\mathcal V_I)^2}{ 3} \Big[\frac 57\Big((M^3)_{\mu \nu}-\frac{1}{10} \tr M^3 g_{\mu \nu}\Big)
-{12}  N^{^{(4125)}}_{\mu\alpha_1\alpha_2,\, \nu}{}^{\beta_1\beta_2}M_{\beta_1}{}^{\alpha_1}M_{\beta_2}{}^{\alpha_2} \Big]\,.
\label{emt-INZ-int4-4}
\eea
For comparison, in $D=6$ the energy-momentum tensor of a non-linear chiral 2-form contains only two terms
\be\label{6DT}
T^{^{(6D)}}_{ \mu  \nu}=\frac 14 M^{^{(6D)}}_{  \mu  \nu}\Big(1-96I_4(\mathcal V^{^{(6D)}}_I)^2\Big)- g_{  \m   \n}\Big(\cV^{^{(6D)}}- 2I_4 \mathcal V^{^{(6D)}}_I \Big)
\ee
which (modulo a normalization) are similar to those in the first line of \eqref{emt-INZ-int4-4}, with the only difference that in $D=6$
\be\label{M6D}
I_4=\tr \left(M^{^{(6D)}}M^{^{(6D)}}\right)\,,\qquad M^{^{(6D)}}_{  \mu  \nu}=\Lambda^{^{(6D)}}_{  \mu  \lambda  \rho}\,\Lambda^{^{(6D)}}_{  \nu}{}^{  \lambda  \rho}\,\qquad (  \mu,  \nu,\ldots=0,1,\ldots,5)
\ee
are built with the components of a self-dual $6D$ 3-form (see Section 3 of \cite{Ferko:2024zth} for details).

In $D=10$ in the second line of \eqref{emt-INZ-int4-4} one can notice the appearance of one of the independent six-order invariants, namely $\tr (M)^3$. Moreover, in $T_{\mu\nu}T^{\mu\nu}$ there appear independent invariants of the 4th, 8th and 12th order in $\Lambda_5$. This difference has drastic consequences for the possibility of interpreting $10D$ non-linear chiral 4-form theories as stress-tensor deformations of the free chiral 4-form theory. As an  illustrative example we will now consider ModMax-type models.

\subsection{ModMax-like conformal duality-invariant theories}\label{MM24610}
In $D=2,4$ and 6 the Lagrangians describing non-linear deformations of free duality-invariant theories that preserve conformal and duality invariance (dubbed ModMax theories) satisfy the following stress-tensor (so-called ``root-$T\bar T$" \footnote{This marginal  $\sqrt{T\bar T}$-operator was first introduced in the context of deformations of two-dimensional CFTs in \cite{Rodriguez:2021tcz} and further explored in \cite{Tempo:2022ndz}.}) flow equation \cite{Babaei-Aghbolagh:2022uij,Ferko:2022iru,Conti:2022egv,Ferko:2022cix}, \cite{Ferko:2024zth}
\begin{align}\label{root_TTbar_flow}
    \frac{\partial \mathcal{L}^{({\gamma)}}}{\partial \gamma} = \frac 1{\sqrt D}\sqrt{ T^{(\gamma)\,\mu\nu} T_{\mu\nu}^{(\gamma)}} \, ,
\end{align}
where $\gamma$ is a dimensionless deformation parameter such that $\mathcal L^{(\gamma=0)}$ is the Lagrangian of the free theory and $T_{\mu\nu}^{(\gamma)}$ is the energy-momentum tensor of the resulting non-linear conformal theory. The operator on the right hand side of \eqref{root_TTbar_flow} is marginal, its dimension is the same as the one  of the Lagrangian density.

To be concrete, let us consider the $6D$ ModMax theory. Its Lagrangian in the INZ-type formulation \cite{Ferko:2024zth} has a form analogous to \eqref{INZ}, but  with $F$ and $\Lambda$ being rank-3 antisymmetric tensors, $E,B$ and $\lambda$ being rank-2 antisymmetric tensors and
\be\label{VMM6}
\mathcal V^{(6D)}(\Lambda_3)=-\frac 1{2\sqrt 6}\tanh\frac \gamma 2 \sqrt{I_4}
\ee
with $I_4$ defined in \eqref{M6D}.

The energy-momentum tensor of the $6D$ ModMax theory obtained from \eqref{6DT} is
\be\label{TMM6}
T_{\mu\nu}^{^{6DMM}}=\frac 14 M^{^{(6D)}}_{\mu\nu}\Big(1-(\tanh\frac \gamma 2 )^2\Big)=\frac 1{4(\cosh\frac\gamma 2)^2} M^{^{(6D)}}_{\mu\nu}\,.
\ee
It is easy to check that the $6D$ Lagrangian with the interaction function \eqref{VMM6} satisfies the flow equation \eqref{root_TTbar_flow}. Vice versa, given the flow equation \eqref{root_TTbar_flow} with the initial condition $\mathcal L^{(0)}=\mathcal L^{free}$, its unique solution is \eqref{VMM6}.
Note that in the limit $\gamma\to \infty$ the energy-momentum tensor of the theory vanishes, so the theory described by the Lagrangian with the function $\mathcal V=-\frac 1{2\sqrt{6}}\sqrt{I_4}$ is trivial \cite{Ferko:2024zth}.

 Let us now have a look at the ten-dimensional analogue of ModMax for which
 \be\label{10DVMM}
 \mathcal V(\Lambda_5)={  b} \sqrt{I_4}\,.
 \ee
First of all, let us note that in $D=4$ and 6 the invariant $I_4$ is always non-negative, as is the term $T_{\mu\nu}T^{\mu\nu}$ (see e.g. item (iv) on page 11 of \cite{Ferko:2024zth} for the proof). In $D=10$, splitting the indices on the time and space ones we have
\be\label{I+}
I_4=M_{\mu\nu}M^{\mu\nu}=(M_{00})^2+M_{ij}M^{ij}-2M_{0i}M_0{}^{i}\,.
\ee
Using Lorentz boosts of $SO(1,9)$ one can choose (at least locally) a reference frame in which the vector $M_{0}{}^i$ is zero. Hence, $I_4$ is also non-negative in $D=10$.

 Substituting the form  \eqref{10DVMM} of the interaction function into  \eqref{emt-INZ-int4-4} we get the $10D$ ModMax energy-momentum tensor
\bea\label{10DMMT1}
T^{^{10DMM}}_{\mu \nu}& =& \frac{1}{2 \cdot 4!} M_{\mu\nu} \Big(1-\frac {24 {  b}^2}{7}\Big)\\
&+&\frac {{  b}^2}{12\,\tr M^2} \Big[\frac 57\Big((M^3)_{\mu \nu}-\frac{1}{10} \tr M^3 g_{\mu \nu}\Big)
-{12}\, (N^{(4125)} MM)_{\mu \nu} \Big]~,\nonumber
\eea
with $(N^{(4125)} MM)_{\mu \nu} = N^{^{(4125)}}_{\mu\alpha_1\alpha_2,\, \nu}{}^{\beta_1\beta_2}M_{\beta_1}{}^{\alpha_1}M_{\beta_2}{}^{\alpha_2}$.
Note that ${  b}^2=\frac {7}{24}$ is a somewhat critical value of the parameter, since the free-theory-looking first term of the energy-momentum tensor vanishes in this case.
So, like in the $6D$ case we can assume that ${  b}^2=\frac 7{24}(\tanh\frac \gamma 2)^2$ with a real parameter $\gamma$.

The square of the energy-momentum tensor \eqref{10DMMT1} is
\bea
T_{\mu \nu} T^{\mu \nu} &=& \frac{1}{4 \cdot (4!)^2} \bigg\{\tr M^2 \Big(1 - \frac{24 {  b}^2}{7}\Big)^2 \non\\
&&+ \frac{8 {  b}^2}{\tr M^2} \Big(1 - \frac{24 {  b}^2}{7}\Big) \Big( \frac{5}{7}\tr M^4  - 12 N^{(4125)}_{\mu \alpha_1 \alpha_2, \nu \beta_1 \beta_2} M^{\mu \nu} M^{\alpha_1 \beta_1} M^{\alpha_2 \beta_2}\Big) \non\\
&&+ \frac{16 {  b}^4}{\big(\tr M^2 \big)^2} \bigg[\frac{25}{49}\tr M^6 -\frac{120}{7} (M^3)^{\mu \nu} N^{(4125)}_{\mu \alpha_1 \alpha_2, \nu \beta_1 \beta_2} M^{\alpha_1 \beta_1} M^{\alpha_2 \beta_2} \non\\
&&\hspace{2cm}-\frac{5}{98} \big(\tr M^3\big)^2 + 144 (N^{(4125)} MM)_{\mu \nu} (N^{(4125)}MM)^{\mu \nu}\bigg] \bigg\}\nonumber\\
&\equiv & \frac{1}{4 \cdot (4!)^2}  \tr M^2 \Big(1 - \frac{24{  b}^2}{7} \Big)^2+\Delta \left(\frac{I_8}{I_4},\frac{I_{12}}{(I_4)^2}\right),
\eea
where
\bea
I_8=\frac 57\tr M^4
-{12} N^{^{(4125)}}_{\mu\alpha_1\alpha_2,}{}^{\nu\beta_1\beta_2}M_\nu{}^{\mu}M_{\beta_1}{}^{\alpha_1}M_{\beta_2}{}^{\alpha_2}
\eea
and
\bea
I_{12} &=& \frac{25}{49}\tr M^6-\frac{5}{98} \big(\tr M^3\big)^2 -\frac{120}{7} (M^3)^{\mu \nu} N^{(4125)}_{\mu \alpha_1 \alpha_2, \nu \beta_1 \beta_2} M^{\alpha_1 \beta_1} M^{\alpha_2 \beta_2}
\nonumber\\
&& + 144 (N^{(4125)} MM)_{\mu \nu} (N^{(4125)}MM)^{\mu \nu}~.
\eea
One can now easily check that the $10D$ ModMax theory with the interaction function \eqref{10DVMM} and $b=-\frac 12 {\sqrt \frac 76}\tanh\frac \gamma 2$ satisfies the flow equation
\be\label{10Droot}
\frac{\partial \mathcal{L}^{({\gamma)}}}{\partial \gamma} = \frac{4!}2\sqrt{ \frac 76}\,\sqrt{ T^{\mu\nu} T_{\mu\nu}-\Delta \left(\frac{I_8}{I_4},\frac{I_{12}}{(I_4)^2}\right)} \, .
\ee
We have thus found that the ``root-$T\bar T$" flow equation \eqref{root_TTbar_flow} which was universal for the ModMax theories in $D=2,4$ and 6 loses its exceptional status in $D=10$. It is modified by higher-order invariants, which are (at least {\it a priori}) not constructed from the components of the energy-momentum tensor.

\subsection{ModMax in the PST formulation}
It is now instructive to have a look at how the $10D$ ModMax Lagrangian looks like in the PST formulation in comparison with the $D=6$ case in which (in the normalization of \cite{Bandos:2020hgy,Ferko:2024zth})
\be\label{MMPST}
\mathcal S^{^{MM}}_{\text{PST}} = \int d^6x \mathcal L^{^{MM}}_{\text{PST}} =\int d^6x\sqrt{-g}\left(\frac 14 E^{ \mu \nu} B_{ \mu \nu}- \cosh(\gamma)s+\sinh(\gamma)\sqrt{s^2-p^2}\right)\,,
\ee
where
\be\label{sp}
 s=\frac 14 B_{ \mu \nu}B^{ \mu \nu},\qquad  p^2=p_{ \mu} p^{ \mu}, \qquad p^{ \mu} =-\frac 1{8\sqrt{-g}}\varepsilon^{ \mu \nu \rho \lambda \sigma \kappa}B_{ \rho \lambda}B_{ \sigma \kappa}v_{ \nu}\,.
\ee
Note that
\be
\sqrt{s^2-p^2}=\frac 1{\sqrt{6}}\sqrt{T_{ \mu \nu}T^{ \mu \nu}}\,\big|_{\gamma=0}\,.
\ee
This action satisfies the stress-tensor flow equation \eqref{root_TTbar_flow} and can be obtained from the INZ-type action
\be\label{INZ6DMM}
\mathcal S^{^{MM}}_{\text{INZ}} =\int d^6x\sqrt{-g}\left(\frac 14 (E^{ \mu \nu} -B^{ \mu \nu})B_{ \mu \nu}+\frac 12(B_{ \mu \nu}+\Lambda_{ \mu \nu \rho}v^\rho)^2+\frac 1{2\sqrt 6}\tanh\frac \gamma 2 \sqrt{I_4(\Lambda_3)}\right)
\ee
upon solving {\it exactly} the equation of motion of $\Lambda_3$ in terms of $B_2$ (see Section 3 of \cite{Ferko:2024zth} for details).

We would like to obtain a PST action for the $10D$ ModMax-like theory from its INZ-type action \eqref{INZ} (with $\mathcal V={  b} \sqrt{I_4(\Lambda)}$) by solving the non-dynamical equation \eqref{Leq}  for $\Lambda$ in terms of $B$ (at least to the first non-linear order) and substituting the result back into the action. To this end we project \eqref{Leq} along $v^{\mu_5}$ and get
\be
\lambda^{\mu(4)}+\frac {5{  b}}{2\sqrt{I_4}}\,\frac{\partial I_4}{\partial\Lambda_{\mu(4)\rho}}v_\rho = -B^{\mu(4)}~.
\label{eom-lambda-int*}
\ee
 To proceed let us use the following identity
which is a consequence of \eqref{Lambda-dec}:
\bea\label{Mdec*}
M_{\m}{}^{\n} =\Lambda_{\mu\rho(4)}\Lambda^{\nu\rho(4)}= \big( \delta_{\m}{}^{\n} + 2 v_{\m} v^{\n}\big)\lambda_{\r(4)}\lambda^{\r(4)} -8\lambda_{\m \r(3)} \lambda^{\n \r(3)} -4 \big(\r_{\m} v^{\n} + \r^{\n} v_{\m} \big)~,
\eea
where
\bea\label{vr}
\r_{\m} = -\frac{1}{4 \cdot 4!} \sqrt{-g} \ve_{\m \r_1 \dots \r_4 \n_1 \dots \n_5} \lambda^{\r_1 \dots \r_4} \lambda^{\n_1 \dots \n_4} v^{\n_5}~.
\eea
With the use of \eqref{Mdec*} and the identity
\bea
\r^2 = \r^{\m}\r_{\m} &=& \frac{1}{8}(\lambda_{\a(4)})^2 (\lambda_{\b(4)})^2 -2\lambda^{\delta\a(3)} \lambda_{\a(3)\g}\lambda^{\g \b(3)} \lambda_{\b(3) \delta}\non\\
&&+ \frac{9}{4} \lambda^{\a(2) \b(2)} \lambda_{\b(2) \g(2)} \lambda^{\g(2) \delta(2)} \lambda_{\delta(2) \a(2)}~,
\eea
the invariant $I_4$ can be written as the sum of $\lambda^4$ invariants as follows
\bea
I_4 (\lambda) = M_{\m}{}^{\n}M_{\n}{}^{\m} &=&
64 (\lambda_{\m \n(3)} \lambda^{\n(3)\s})( \lambda_{\s \rho(3)} \lambda^{\rho(3) \m}) -6 (\lambda_{\a(4)})^2 (\lambda_{\b(4)})^2- 32 \rho^2 \nonumber \\
&=& -10 (\lambda_{\a(4)})^2 (\lambda_{\b(4)})^2 + 128 \lambda^{\delta\a(3)} \lambda_{\a(3)\g}\lambda^{\g \b(3)} \lambda_{\b(3) \delta} \non\\
&&-72 \lambda^{\a(2) \b(2)}\lambda_{\b(2) \g(2)} \lambda^{\g(2) \delta(2)} \lambda_{\delta(2) \a(2)}~.
\label{I-lambda}
\eea
Then, the equation \eqref{eom-lambda-int*} takes the following form
\be
\lambda^{\mu(4)}- \frac {  b} {4\sqrt{I_4}} \,\frac{\partial I_4(\lambda)}{\partial\lambda_{\m(4)}} = -B^{\mu(4)}~.
\label{eom-lambda-int1}
\ee
We now solve this equation for $\lambda$ perturbatively up to the first order in ${  b}$.

At the zeroth order in ${  b}$, using \eqref{Lambda-dec} we have
\be\label{L0}
\Lambda_{\mu_1\mu_2\mu_3\mu_4\mu_5} = -5 B_{[\m_1 \dots \m_4} v_{\m_5]}-\frac{1}{4!}\sqrt{-g} \ve_{\m_1 \dots \m_5 \n_1 \dots \n_5}B^{\n_1 \dots \n_4} v^{\n_5}~,
\ee
and at the linear order in $b$
\be\label{L1}
\lambda^{\mu(4)}
=-B^{\mu(4)}-\frac {  b} {4\sqrt{I_4(B)}} \frac{\partial I_4(B)}{\partial B_{\m(4)}}~,
\ee
 where
\be\label{I-B}
I_4(B)  =
64 (B_{\m \n(3)} B^{\n(3)\s})( B_{\s \rho(3)} B^{\rho(3) \m}) -6 (B^2)^2- 32 p^2
\ee
 is the quartic invariant originally found in the PST formulation in \cite{Buratti:2019guq}.
Note also that
\be\label{I4PST=root TT}
I_4(B)
=(2\cdot 4!)^2\,T_{\mu\nu}T^{\mu\nu}\vert_{\rm PST,free}\,,
\ee
where $T^{\rm PST,free}_{\mu\nu}$ was given in \eqref{emt-pst-free}\,.

Substituting the expression \eqref{L1} for $\lambda$ into the action \eqref{INZ} with $\mathcal V=b\sqrt{I_4}$  we get (up to the second order in ${  b}^2$)
\bea
\mathcal S
= \frac{1}{4!} \int d^{10}x \sqrt{-g} \,\Big( \hf E \cdot B- \hf B \cdot B-{  b} \sqrt{I_4(B)} -\frac{{  b}^2}{16}\frac {I_6(B)}{I_4(B)}\,\Big)\,,
\label{PSTMM10}
\eea
where
\be\label{I6B}
I_6(B)=\left(\frac {\partial I_4}{\partial B}\cdot \frac{\partial I_4}{\partial B}\right)
\ee
is a 6th-order invariant in $B$ obtained by taking the square of the $B$-derivative of the 4th-order invariant \eqref{I-B}. We do not give its explicit form because it is not suggestive.

We see that in contrast to the $6D$ case \eqref{MMPST}, at the second order in the parameter $b$ the ten-dimensional PST action for ModMax acquires a term that involves invariants of the sixth order in the field $B_4$. At higher orders in $b$ there will appear higher order invariants of $B_4$. It is not clear whether the complete action can be written in the closed form. This means, in particular, that the physical Hamiltonian density of the $10D$ ModMax-like theory based on the INZ interacting function $\mathcal V={  b}\sqrt{I_4}$ is very complicated. This Hamiltonian density is obtained from the PST action in which the auxiliary vector $v_\mu$ is gauge fixed along the time direction. For instance, in flat space-time we can fix $v_\mu=\delta_\mu^0$. Then the Hamiltonian density is
\be\label{Ham}
\mathcal H^{^{10DMM}}(B)=\frac {1}{4!}\Big(\frac 12 (B\cdot B)+{  b}\sqrt{I_4(B)} +\frac{{  b}^2}{16}\frac {I_6(B)}{I_4(B)}+O(b^3)\Big)\,,
\ee
where now $B^{ijkl}(x)$ has only spatial indices ($i,j,..=1,\ldots,9)$ and is the genuine physical magnetic-like field of the theory. The Lagrangian of such a theory can hardly satisfy the root-$T\bar T$ flow equation \eqref{root_TTbar_flow}.

\section{INZ versus PST approach in $4n+2$ dimensions} \label{INZvsPST}
In this Section we elaborate and refine the proof of the classical equivalence of the INZ- and PST-like formulations of non-linear $D=4n+2$ chiral $2n$-form theories given in \cite{Ferko:2024zth}. The only theory for which the INZ counterpart of the PST formulation does not exist \cite{Ferko:2024zth} is the non-linear conformal chiral $2n$-form theory of the Bia\l{}ynicki-Birula type \cite{Bialynicki-Birula:1984daz,Bialynicki-Birula:1992rcm,Gibbons:2000ck,Townsend:2019ils,Bandos:2020hgy}. We will give the form of the PST action for this theory in Section \ref{BB}.

\subsection{From INZ to PST and back}
In $D=4n+2$ dimensions we consider a theory describing the dynamics of a chiral gauge field characterized by an antisymmetric potential $A$ of rank $2n$ and its curl $F=dA$ of rank $2n+1$, such that on the mass shell $F$ satisfies a non-linear self-duality condition similar to \eqref{F^-=VF+} in $D=10$. To construct a manifestly space-time invariant action of this theory we use the PST vector $v^\mu=\partial^\mu
a/{\sqrt{-(\partial a)^2}}$ and introduce the electric and magnetic fields $E=i_vF$ and $B=i_v \widetilde F$. In the INZ-type formulation of the theory we also introduce a self-dual rank-$(2n+1)$ auxiliary field $\Lambda=\widetilde \Lambda$. We denote the contraction of $\Lambda$ with $v$ as $\lambda=i_v\Lambda$. In the INZ approach the action is then given by (here we have changed $\lambda$ to $-\lambda$, so that in the free case we have $\lambda=B$, with the plus sign)
\be\label{inzd}
I[A, a, \L] = \frac{1}{(2n)!} \int d^Dx \sqrt{-g}\, \Big( \frac{1}{2} \,E \cdot B- \frac{1}{2}\, B\cdot  B +
(\lambda+B)\cdot (\lambda+B) - \cV(\Lambda)\Big),
\ee
where $\cV(\Lambda)$ is the Lorentz-invariant interaction potential of the auxiliary field. The action \eqref{inzd} is invariant under the PST-like transformations
\be\label{PST11}
\delta A_{2n}=v\wedge \Phi_{2n-1}(x),\quad \delta a=0,\quad \delta \Lambda=0,
\ee
\bea\label{symmd}
\delta a=\varphi(x), \quad \delta A=- \frac{\varphi}{\sqrt{-(\partial a)^2}}\,(E+B+2\lambda),\quad \delta\Lambda=0.
\eea
For a generic $2n+1$-form we have the general decomposition
\be\label{fdecomp}
F^{\mu_1\cdots\mu_{2n+1}}=-(2n+1)E^{[\mu_1\cdots\mu_{2n}}v^{\mu_{2n+1}]}-\frac{1}{(2n)!}\,\varepsilon^{\mu_1\cdots\mu_{2n+1}\nu_1\cdots\nu_{2n+1}}
B_{\nu_1\cdots\nu_{2n}}v_{\nu_{2n+1}}\,,
\ee
which for a self-dual $2n+1$-form is
\be\label{selfdual}
\Lambda^{\mu_1\cdots\mu_{2n+1}}=-(2n+1)\lambda^{[\mu_1\cdots\mu_{2n}}v^{\mu_{2n+1}]}-\frac{1}{(2n)!}\,\varepsilon^{\mu_1\cdots\mu_{2n+1}\nu_1\cdots\nu_{2n+1}}
\lambda_{\nu_1\cdots\nu_{2n}}v_{\nu_{2n+1}}\,.
\ee
This allows us to trade the interaction term $\cV(\Lambda)$ as a function of $\lambda$,  which we still denote by $\cV(\lambda)$. We may then introduce the derivatives\footnote{We consider the derivatives of a scalar function w.r.t. an antisymmetric tensor field normalized according to the convention $\delta \cV=(\partial \cV/\partial\Lambda_{\mu_1\cdots \mu_{2n+1}})\,\delta\Lambda_{\mu_1\cdots \mu_{2n+1}}$.}
\be\label{mmn}
W^{\mu_1\cdots \mu_{2n+1}}=\frac{\partial \cV}{\partial \Lambda_{\mu_1\cdots \mu_{2n+1}}},\quad\quad w^{\mu_1\cdots \mu_{2n}}=\frac{\partial \cV}{\partial \lambda_{\mu_1\cdots \mu_{2n}}}.
\ee
Notice that, since $\Lambda$ is self-dual, $W$ is an anti-self-dual, $\widetilde W=-W$. Since $\lambda=i_v\Lambda$, we have the relation
\be\label{mM}
w=-D\,i_vW=D\,i_v\widetilde W,
\ee
and the identity
\be\label{mlml}
(W\cdot\Lambda)=(w\cdot\lambda)\,.
\ee
We can now eliminate from \eqref{inzd} the auxiliary field $\L$ via its equation of motion
\be\label{auxd}
\lambda+B=\frac w2.
\ee
This equation can be used to express $\lambda$ in terms of $B$, obtaining the function $\lambda(B)$, at least order-by-order in perturbation theory.\footnote{ In general, it is difficult to solve equation \eqref{auxd} for $\lambda$ explicitly, especially in  $D=10$ and higher dimensions. However, in $D=6$ a few examples of explicit solutions exist, including chiral 2-form ModMax and Born-Infeld theories (see \cite{Ferko:2024zth} for details). It would be interesting to find an example of a $D=10$ chiral 4-form theory for which the solution of \eqref{auxd} exists in a closed form.}
Replacing this function in \eqref{inzd} we obtain a PST-like action
\be\label{pstd}
I[A,a] = \frac{1}{(2n)!} \int d^Dx \sqrt{-g}\, \Big( \frac{1}{2} \,E  B-  \mathcal H(B)\Big),
\ee
where the PST interaction function is given by
\be\label{hd}
{\cal H}(B)= \cV(\Lambda)+ \frac{1}{2}\, BB -
(\lambda+B) (\lambda+B) \bigg|_{\lambda=\lambda(B)}.
\ee
Thanks to \eqref{auxd} we can express the derivative
\be\label{hdb}
H^{\mu_1\cdots \mu_{2n}}=\frac{\partial {\cal H}}{\partial B_{\mu_1\cdots \mu_{2n}}}
\ee
in the form
\be\label{hlb}
H=-2\lambda-B.
\ee
The above relation allows one to explicitly reconstruct $\Lambda$ (with the use of \eqref{selfdual}) as a function of $B$ and a given $\mathcal H(B)$.

On general grounds, once we have eliminated the auxiliary field $\L$ via its algebraic equation of motion \eqref{auxd}, the action \eqref{pstd} becomes automatically invariant under the transformations \eqref{symmd}, once we have replaced in these transformations $\lambda$ with
$\lambda(B)$. Thanks to the relation \eqref{hlb} the resulting transformation laws read
\bea
\delta a=\varphi, \quad \delta A=- \frac{\varphi}{\sqrt{-(\partial a)^2}}\,(E-H),
\eea
which are precisely the symmetries of the PST formulation. Furthermore, we know that these symmetries of the action \eqref{pstd} require that the function $\mathcal H(B)$ satisfies the following  condition
\be\label{pstcd}
\ve^{\alpha\beta\mu_1\cdots \mu_{2n}\nu_1\cdots \nu_{2n}}
(B_{\mu_1\cdots \mu_{2n}} B_{\nu_1\cdots \nu_{2n}}-
H_{\mu_1\cdots \mu_{2n}} H_{\nu_1\cdots \nu_{2n}})v_\beta=0.
\ee
We can thus conclude that the INZ approach gives rise to an action in the PST formulation which satisfies automatically the PST condition \eqref{pstcd}.

We provide now a direct proof of this statement, which will show in particular that also the converse is true, namely, with the exception of the Bia\l{}ynicki-Birula theory, for each action in the PST approach there exists an action \eqref{inzd} of the INZ approach in which the function $\cV(\L)$ is Lorentz invariant and $v$-independent. Starting from the INZ approach, we must prove the validity of the PST condition \eqref{pstcd}. To this end, using \eqref{auxd} and \eqref{hlb}, we put the latter in the equivalent form
\be\label{PSTcon2}
\ve^{\alpha\beta\mu_1\cdots \mu_{2n}\nu_1\cdots \nu_{2n}}
\lambda_{\mu_1\cdots \mu_{2n}}w_{\nu_1\cdots \nu_{2n}}v_\beta=0.
\ee
This equation has been first derived in $D=6$ in \cite{Ferko:2024zth}. Next, in this equation we replace  $w$ with $i_v\widetilde W$, see \eqref{mM}, and express the product of the Levi-Civita tensors in terms of Kronecker $\delta$'s. With the use of \eqref{mlml} we can thus reduce \eqref{PSTcon2} to
\be\label{PSTcon3}
D(W\lambda)^\alpha-v^\alpha(W\cdot\Lambda)=0,
\ee
where the contracted indices have not been written explicitly. This relation can also be written as
\be\label{intd}
v_\beta(W\Lambda)^{\alpha\beta}-\frac 1D\,v^\alpha(W\cdot\Lambda)=0.
\ee
The last identity we need is
\be
(W\L)^{(\alpha\beta)}=\frac{1}{D}\,(W\cdot\Lambda)\,\eta^{\alpha\beta},
\ee
which can be proven considering that $(W\L)^{\alpha\beta}=-(\widetilde W\,\widetilde\L)^{\alpha\beta}$, and working out the product of Levi-Civita tensors on the r.h.s. With this identity, the condition \eqref{intd} can be rewritten as
\be\label{vml}
v_\beta(W\L)^{[\alpha\beta]} =0,
\ee
which is thus equivalent to the PST condition \eqref{pstcd}. We can now resort to the Lorentz-invariance of the INZ interaction function $\cV(\L)$. For an infinitesimal Lorentz transformation with an antisymmetric transformation parameter $l_{\alpha \beta}$
we have
\bea
\delta_l \cV(\L)=\delta_l \L \frac{\partial \cV}{\partial\L}=   \delta_l \Lambda W=(2n+1)\,l_{\alpha\beta}(W\L)^{\alpha\beta}=0.
\eea
As this identity holds for all antisymmetric $l_{\alpha\beta}$, we obtain
\bea
(W\L)^{[\alpha\beta]}=0.
\eea
The condition \eqref{vml} is thus a consequence of the Lorentz invariance of $\cV(\L)$.

Conversely, let us start from a PST action \eqref{pstd} with an interaction function ${\cal H}(B)$ satisfying the condition \eqref{pstcd}. Via the relation \eqref{hlb} we can determine the rank-$2n$ field $\lambda(B)$ as a function of $B$. Equation \eqref{hd} then determines the Lorentz-invariant INZ interaction function $\cV(\lambda)$ as a function of $\lambda$. Furthermore, equation \eqref{selfdual} allows to construct the self-dual tensor $\Lambda$ from $\lambda$. At this point the INZ interaction function is actually a Lorentz-invariant function of $\L$ and $v$, i.e. $\cV(\lambda)\equiv\cV(\L,v)$. Note that the PST condition \eqref{pstcd} is still equivalent to \eqref{PSTcon2}. We will now show that if $w=\partial \mathcal V/\partial \lambda$ satisfies the PST condition \eqref{PSTcon2}, then $\mathcal V(\L, v)$ actually does not depend on $v$, i.e. $\mathcal V(\L,v)=\cV(\L)$.
To this end we recall that $v^\mu$ is a composite unit vector of the form $v^\mu=u^\mu/\sqrt{-u^2}$ ($u_\mu=\partial_\mu a(x)$). Therefore, we have
\be
\frac{\partial \cV}{\partial u^\mu}=\frac{1}{\sqrt{-u^2}}\left(\eta_{\mu\nu}+v_\mu v_\nu\right) \frac{\partial \cV}{\partial v_\nu}.
\ee
Using the decomposition \eqref{selfdual} of the tensor $\Lambda$ we find
\be
\frac{\partial \cV}{\partial v_\nu}=(\Lambda w)^\nu=-(\lambda w)v^\nu
-\frac{1}{(2n)!}\,\varepsilon^{\nu\mu_1\cdots\mu_{2n}\nu_1\cdots\nu_{2n}\rho}
\lambda_{\nu_1\cdots\nu_{2n}}w_{\mu_1\cdots\mu_{2n}}v_\rho.
\ee
The last term in this relation vanishes thanks to \eqref{PSTcon2}, and so we get $\partial \cV/\partial u^\mu=0$. We have thus shown that, if the PST interaction function ${\cal H}(B)$ satisfies the condition \eqref{pstcd}, then the INZ Lorentz-invariant interaction function $\cV(\Lambda,v)$ does not depend on $v$.
This also implies that the non-linear self-duality equation of the PST approach, namely $E_{2n}=H_{2n}(B)$, can (almost) always be recast in a manifestly Lorentz invariant way that does no longer involve $v$, see eq. \eqref{F^-=VF+} in $D=10$. The only theory that does not fit into this general consideration is the Bia\l{}ynicki-Birula theory.

\subsection{Bia\l{}ynicki-Birula theory}\label{BB}
In $D=4$ this specific non-linear conformal duality-invariant electrodynamics theory was found by Bia\l{}ynicki-Birula  \cite{Bialynicki-Birula:1984daz,Bialynicki-Birula:1992rcm} by taking a strong field limit of the Born-Infeld theory in its Hamiltonian formulation. The four-dimensional Bia\l{}ynicki-Birula theory exhibits the enhancement of the electric-magnetic duality group from $U(1)$ to $SL(2,R)$ and has an infinite number of conserved currents. Analogous theories were later found as strong field limits of non-conformal six and higher-dimensional non-linear $p$-form theories \cite{Gibbons:2000ck,Townsend:2019ils,Bandos:2020hgy}.

The Bia\l{}ynicki-Birula (BB) electrodynamics is the only theory within the known variety of duality-invariant non-linear $p$-form theories whose action is known only in the first-order Hamiltonian form, or equivalently in the covariant PST formulation. The non-linear (Hamiltonian) density $\mathcal H(B)$ of the $D=(4n+2)$ BB theory in the action \eqref{pstd} is
\be\label{BBH}
\mathcal H_{BB}=2\sqrt{p_\mu p^\mu}\,,
\ee
where
\be \label{p-dimD}
p^\mu=-\frac 1{4\cdot (2n)!\sqrt{-g}}\varepsilon^{\mu\nu(2n)\rho(2n)\lambda}B_{\nu(2n)}B_{\rho(2n)}v_\lambda=-\frac{1}{4} \tilde{B}^{\mu \nu(2n)} B_{\nu(2n)}\,,
\ee
with
\be\label{tildeB}
\tilde B_{\mu(2n+1)}\equiv \frac {\sqrt{-g}}{(2n)!}\varepsilon_{\mu(2n+1)\kappa(2n)\sigma}B^{\kappa(2n)}v^\sigma\,.
\ee
It can be checked that this $\mathcal H(B)$ satisfies the PST gauge invariance condition \eqref{pstcd}.

We will now show that it is not possible to reconstruct the INZ-type formulation of the Bia\l{}ynicki-Birula theory from its PST action. To pass from PST to INZ one should invert the equation \eqref{hlb}, i.e. to compute $B$ as a function of $\lambda$ and substitute the result into \eqref{hd} to get $\mathcal V(\lambda)$. In the attempt to solve \eqref{hlb} for $B$ in the case of \eqref{BBH} we notice that
\be\label{H=ntB}
H_{\mu(2n)}=\frac{\partial\mathcal H}{\partial B^{\mu(2n)}}=-\frac{p^\nu}{\sqrt{p^2}}\tilde B_{\nu\mu(2n)}\,,
\ee
or
\be\label{H=ntB1}
H_{\mu(2n)}=-n^\nu\tilde B_{\nu\mu(2n)}\,\qquad n^\mu=\frac{p^\mu}{\sqrt{p^2}}\,,
\qquad n^\mu n_\mu=1\,.
\ee
Now, for simplicity, we set $v_\mu=\delta_\mu^0$, obtaining  ${\cal H}_{BB}=2\sqrt{p^jp^j}$ ($j=1,\ldots, 4n+1$), and
\be
H^{j_1\cdots j_{2n}}=- n^i\widetilde{B}^{ij_1\cdots j_{2n}},\quad {\rm with}\quad  n^i=\frac{p^i}{\sqrt{p^jp^j}},
\ee
where $\widetilde B$ is the $(4n+1)$-dimensional Hodge-dual of $B^{j_1\cdots j_{2n}}$.
So equation \eqref{hlb} becomes
\be\label{bbn}
B^{j_1\cdots j_{2n}} - n^i\widetilde{B}^{ij_1\cdots j_{2n}}=-2\lambda^{j_1\cdots j_{2n}}.
\ee
If we decompose $B$ in its components parallel and orthogonal   to $n^i$, $B=B_{\parallel}+B_\perp$, we then see that \eqref{bbn} determines $B_\parallel$ in terms of $\lambda$, while it determines only the ($4n$-dimensional) antiself-dual part of $B_\perp$ in terms of $\lambda$, leaving its self-dual part undetermined. We thus conclude that for the Hamiltonian \eqref{BBH} the relation \eqref{hlb} cannot be inverted to compute $B$ as a function of $\lambda$, and so the INZ formulation of the BB theory cannot be obtained.

A consequence of this is that a non-linear duality relation of the form
\be\label{durs}
F^-_{2n+1}=\frac 12 (F_{2n+1}-\tilde F_{2n+1})=-(2n+1)\frac{\partial \cV(F^+)}{\partial F^+_{2n+1}}
\ee
is not known for the BB theory. A different covariant form of the $6D$ BB chiral 2-form equations can be obtained by taking a tensionless limit of the chiral 2-form part of the M5-brane equations of motion in the superembedding approach \cite{Gibbons:2000ck}.

 One of the reasons of the difficulty to obtain other (more conventional) Lagrangian formulations of the BB theory is that, e.g. in $D=4$ it describes the dynamics of only those electromagnetic fields which are null, i.e. whose field strengths satisfy the conditions $F_{\mu\nu}F^{\mu\nu}=F_{\mu\nu}\tilde F^{\mu\nu}=0$. Analogously, in $D=6$ the BB rank-3 field strength (which is not Hodge self-dual) is null $F_{\mu\nu\rho}F^{\mu\nu\rho}=0$ together with the 4th order invariant $I_4=(F^+)^4=0$.  Also $T_{\mu\nu}T^{\mu\nu}=0$ for the BB theories in $D=4,6$ \cite{Ferko:2024zth,Deger:2024jfh,Russo:2025wph}. Moreover, we will show that $T_{\mu\rho}T^{\rho\nu}=0$ for the BB theories in any dimension. These conditions arise as consequences of the equations of motion of the Hamiltonian (or PST) formulation of the BB theory. In a `conventional' Lagrangian formulation one would need to introduce the null-field conditions with Lagrange multipliers in the action, as was suggested in \cite{Bialynicki-Birula:1992rcm}, but it is not obvious how to decouple these Lagrange multipliers from the physical field strengths in the equations of motion.

In $D=10$ and higher dimensions the BB theory does not describe null fields with \linebreak\hbox{$F_{\mu(2n+1)}F^{\mu(2n+1)}=0$.} Indeed, let us compute  $F_{\mu(2n+1)}F^{\mu(2n+1)}$ using the decomposition \eqref{fdecomp} and the self-duality equation  of motion $E_{2n}=H_{2n}$ in the PST formulation. We thus get
$$F^2=-(2n+1)( E^2-B^2)=-(2n+1)(H^2-B^2),$$
and since
\be\label{H2BB}
H^2=(n_\mu\widetilde{B}^{\mu\nu(2n)})^2=B^2-(2n)(n^\mu B_{\mu\nu(2n-1)})^2\,, \qquad  \ee
we have
\be\label{F_D/2^2}
F_{\mu(2n+1)}F^{\mu(2n+1)}=2n(2n+1)(n^\mu B_{\mu\nu(2n-1)})^2\not =0,\quad for\quad n>1.
\ee
Note that in $D=6$ ($n=1$) the following identity holds $n^\mu B_{\mu\nu}\equiv 0$ and hence $FF=0$.

Let us now have a look at the energy-momentum tensor of the BB theory in $D=4n+2$.
The energy-momentum tensor of the PST formulation derived from \eqref{pstd} has the following form
\be\label{Tgen}
T_{\mu\nu}= \frac 1{(2n)!}\Big(v_{\m} v_{\n} H_{\r(2n)}B^{\r(2n)} - 2n \,H_{(\m}{}^{\r(2n-1)} B_{\n) \r(2n-1)}- g_{\m \n} \big( \cH - \frac 12 H_{\r(2n)} B^{\r(2n)}\big) -4p_{(\m} v_{\n)} \Big)~.
\ee
For the BB theory with $\mathcal H$ defined in \eqref{BBH} and $H_{2n}$ in \eqref{H=ntB}
\be\label{BBemt-D}
T_{\mu\nu}^{BB}= \frac 1{(2n)!}\Big(2(g_{\m \n} +2 v_{\m} v_{\n})\sqrt{p_\mu p^\mu} + 2n\, n^\rho \tilde B_{\rho(\m}{}^{\lambda(2n-1)} B_{\n) \lambda(2n-1)}-4p_{(\m} v_{\n)}\Big)~,
\ee
Using the identity
\be\label{identity-D}
B^{\rho}{}_{\alpha(2n-1)}\tilde B^{\mu\nu\alpha(2n-1)}=\frac 1n (\eta^{\rho\mu} p^\nu-\eta^{\rho\nu}p^\mu)+ \frac 1n v^\rho (v^\mu p^\nu-v^\nu p^\mu).
\ee
 we  obtain
\bea
T_{\mu\nu}^{BB}=\frac 1{(2n)!}\bigg(2 v_{\m} v_{\n}\sqrt{p^2} + 2 \frac{p_{\mu} p_{\nu}}{\sqrt{p^2}} -4p_{(\m} v_{\n)}\bigg)=\frac {2\sqrt{p^2}}{(2n)!}(v_\mu-{n_\mu})(v_\nu-{n_\nu})~.
\eea
From this expression it follows that in the BB theory $T_{\mu\rho}T^{\rho\nu}=0$. As such this theory cannot be realized as a $T\overline T$-like deformation of a seed chiral $2n$-form theory in $D=4n+2$.

Finally, let us note that, as in the $D=6$ case \cite{Ferko:2024zth}, on the mass shell the BB energy-momentum tensor can be written in a manifestly covariant form without the use of the auxiliary vector $v_\mu$. To this end one should use the equation of motion of the gauge potential $A_{\mu(2n)}$, which produces the non-linear self-duality condition
$$
E_{\mu(2n)}=H_{\mu(2n)}\,.
$$
One then replaces $H_{\mu(2n)}$ with $E_{\mu(2n)}$ in \eqref{Tgen}, notices that  on the mass shell (due to the conformal invariance) $\mathcal H_{BB}-\frac 12 E\cdot B=0$ and thus obtains
\bea
T^{BB}_{\mu\nu}&= &\frac 1{(2n)!}\Big(v_{\m} v_{\n} E_{\r(2n)}B^{\r(2n)} - 2n \,E_{(\m}{}^{\r(2n-1)} B_{\n) \r(2n-1)} -4p_{(\m} v_{\n)} \Big)\nonumber\\
&=&\frac 1{2(2n)!}F_{\rho(2n)(\mu}\tilde F_{\nu)}{}^{\rho(2n)}\,,
\eea
where the last expression arises upon the use of the identity \eqref{fdecomp}.

\section{Equivalence of  INZ and ``clone" formulations of non-linear duality-invariant $p$-form theories } \label{INZvsclone}
We shall now show that a so-called ``clone" Lagrangian formulation of non-linear duality-invariant $p$-form theories introduced for free theories by Mkrtchyan \cite{Mkrtchyan:2019opf} and recently generalized to non-linear cases in \cite{Avetisyan:2021heg,Avetisyan:2022zza} is fully equivalent  { to the INZ formulation \cite{Ivanov:2014nya} originally put forward for $D=4$ non-linear electrodynamics and generalized recently to higher-dimensional duality-invariant $p$-form theories in \cite{Ferko:2024zth}}.

We will demonstrate this equivalence for the case of the non-linear chiral $2n$-form theories in $D=4n+2$, using a simple field redefinition. The generalization of this proof to duality-invariant $(2n+1)$-form theories in $D=4(n+1)$ is straightforward. It can be obtained, for instance, by performing the dimensional reduction and truncation of the $D=4n+2$ theory to $D=4n$.

The non-linear chiral $2n$-form ``clone" Lagrangian density has the following form in our conventions
\be\label{clone}
\mathcal L=-\frac 14\Big((\hat F+a(x)Q)\cdot (\hat F+a(x)Q)- 2a(x) Q\cdot \widetilde {\hat F}\Big)-\mathcal V(\Lambda),
\ee
where $\hat F_{2n+1}=d\hat A_{2n}$, $Q_{2n+1}=d R_{2n}$, $a(x)$ is a PST-like scalar and
\be\label{Lclone}
\Lambda\equiv -\frac 12 \Big(\hat F+\widetilde {\hat F}+a(x)(Q+\tilde Q)\Big)
\ee
is the self-dual part of $\hat F+aQ$. The Lagrangian possesses a number of local symmetries whose detailed description is given e.g. in \cite{Avetisyan:2022zza}. We will only need one of them for our purposes as will be explained below. { At this point we should only note that local symmetry transformations in the ``clone" formulation contain the factor $\frac 1{(\partial a)^2}$ (similar to the one in  \eqref{symmd}). Hence, the peculiarity that $\partial_\mu a(x)$ must be a nowhere vanishing vector remains also in this formulation (see e.g. Section 4 of \cite{Evnin:2022kqn} for the discussion of this point). }

The (2n+1)-form $\hat F+a(x)Q$ can be redefined as follows
\be\label{F+aQ}
\hat F+a(x)Q=F-da\wedge R,
\ee
where
$$
F_{2n+1}=d A_{2n}=\hat F_{2n+1}+d(aR_{2n})\,.
$$
Then,
up to a total derivative the Lagrangian density is equivalently rewritten as follows
\be\label{clone1}
\mathcal L=-\frac 14\Big((F-(i_{da}\tilde R)^* )\cdot (F-(i_{da}\tilde R)^*)+ 2 R\cdot (i_{da}\tilde F)\Big)-\mathcal V(\Lambda),
\ee
where $^*$ stands for the Hodge dual of $i_{da}\tilde R$.\footnote{\label{KM} Let us note that the free part of the Lagrangian density \eqref{clone1}, with $\partial_\mu a$ replaced by an arbitrary auxiliary vector field $c_\mu$ and the Lagrange multiplier term $G^{[\mu\nu]}\partial_\mu c_\nu$ added to the action, was the starting point of \cite{Mkrtchyan:2019opf}. The equation of motion of the Lagrange multiplier $G^{\mu\nu}$ sets $c_\mu=\partial_\mu a$. The free term in \eqref{clone} was then obtained in \cite{Mkrtchyan:2019opf} by the field redefinition inverse to \eqref{F+aQ} performed in \eqref{clone1} (with $\mathcal V(\Lambda)=0$).}

Note that the components of $ R$ along $\partial_\mu a(x)$, i.e. $da\wedge i_{da} R$, do not contribute to the Lagrangian density.  This is because $R$ appears always in the combination $da \wedge R$ (see eq. \eqref{F+aQ}), or equivalently as $i_{da}\tilde R$ in \eqref{clone1}. Hence, without  loss of generality we can consider $ R$ such that $i_{da} R=0$. This reflects the presence of the symmetry similar to \eqref{PST11}.

{Now note that $\Lambda$ in \eqref{Lclone} takes the form
\be\label{Lclone1}
\Lambda=-\frac 12 \Big(  F+\tilde {  F}-da\wedge R-(da\wedge R)^*\Big)\,.
\ee
\if{}
where the self-dual (2n+1)-form $Y$ is defined in accordance with the identity \eqref{selfdual} such that
$$Y=\tilde Y=-\Big(v\wedge R+(v\wedge R)^*\Big)\,, \qquad i_vY=R,\qquad v=\frac {da}{\sqrt{-(\partial a)^2}}\,.$$
\fi
At this point in the Lagrangian density \eqref{clone1} we have the auxiliary anti-symmetric tensor field $R_{2n}$, while the composite self-dual tensor field $\Lambda_{2n+1}$ is  related to $R_{2n}$ via \eqref{Lclone1}. $R_{2n}$ and $\Lambda$ have the {\it same number} of independent components.
Therefore, instead of $R_{2n}$ we can take $\Lambda$ as the independent auxiliary self-dual field.\footnote{ This field redefinition of the auxiliary field $R_{2n}$ is legitimate though it involves derivatives of other fields (one of which is the auxiliary field $a(x)$). A reason is that this is a redefinition of a {\it non-dynamical} field, it does not change the number of dynamical degrees of freedom and the physical field equations of motion of the theory (indeed, the "clone" and the INZ actions produce exactly the same physical field equation \eqref{durs}). An alternative way to convince oneself of the consistency of this field redefinition is to remove the derivatives of the fields by replacing $\partial_\mu a$ with an auxiliary vector field $c_\mu$ and adding the Lagrange multiplier term $G^{[\mu\nu]}\partial_\mu c_\nu$ to the action (as was mentioned in the footnote \ref{KM}), and to perform the same operation for $F_{2n+1}=dA_{2n}$. Then one makes the algebraic field redefinition $R_{2n}\to \Lambda$ with the use of these new auxiliary fields and in the end one solves for the Lagrange multiplier equations of motion in the final action composed of $A_{2n}, a(x)$ and $\Lambda=\tilde \Lambda$.}
Then taking the inner product of $v=d a/{\sqrt{-\partial a^2}}$ with $\Lambda$ in eq. \eqref{Lclone1},  we have
\be\label{lambda=}
\lambda_{2n}\equiv(i_v\Lambda)_{2n}=-\frac 12 (E_{2n}+B_{2n}+{\sqrt{-(\partial a)^2}} R_{2n}),
\ee
where (as in the previous Sections)
$$
E_{2n}=(i_vF)_{2n}\,,\qquad B_{2n}=(i_v\tilde F)_{2n}
$$
and hence
\be\label{R}
R_{2n}=-\frac 1{\sqrt{-(\partial a)^2}}\,(2\lambda_{2n}+E_{2n}+B_{2n})\,.
\ee
Substituting \eqref{R} into \eqref{clone1} and upon a straightforward algebraic rearrangement of terms, we arrive at
\bea\label{clone3}
\mathcal L&=&
\if{}
-\frac 14\Big(-\Lambda\cdot (F-\tilde F)-4\lambda^2-2(E+B)\cdot \lambda -4\lambda\cdot B-2(E+B)\cdot B\Big)-\mathcal V(\Lambda)\nonumber\\
&=&-\frac 14\Big(2\lambda\cdot (E-B)-4\lambda^2-2(E-B)\cdot \lambda -8\lambda\cdot B-2(E+B)\cdot B\Big)-\mathcal V(\Lambda)\nonumber\\
&=&
\fi
\frac 12(E+B)\cdot B+2\lambda\cdot B+\lambda^2-\mathcal V(\Lambda).
\eea
The above Lagrangian density coincides with the INZ one \eqref{inzd}. Vice versa, by performing the inverse field redefinition one can derive the ``clone" Lagrangian density \eqref{clone} from the INZ one \eqref{clone3}.

The following final comment is in order. The ``clone" Lagrangian density \eqref{clone} appears polynomial in the fields (if the interaction function $\mathcal V(\Lambda)$ is polynomial) while the INZ (and PST) actions are non-polynomial, since they contain the factors of $\frac 1{(\partial a)^2}$ entering via the composite vector field $v=d a/\sqrt{-(\partial a)^2}$. But, as we have already pointed out, in spite of the apparent polynomiality of the ``clone" action the non-polynomial factor $\frac 1{(\partial a)^2}$ is hidden in local symmetry transformations of fields of this formulation.

The PST and INZ actions can also be made to look polynomial by introducing additional auxiliary fields and Lagrange multiplier terms similar to those used in the original papers \cite{Pasti:1995ii,Pasti:1995tn}. For instance, one can treat $v_\mu$ as an independent unit time-like ($v^2=-1$) auxiliary vector field (a kind of einbein similar to the vielbein in the Cartan formulation of gravity). Then to relate this $v_\mu$ to $\partial_\mu a$ one can add to the PST or INZ action the Lagrange multiplier term $c^\mu(v_\mu-c \partial_\mu a)$ with $c_\mu$ and $c$ being auxiliary fields. Also the condition $v^2=-1$ can be included into the action with a Lagrange multiplier. It is an easy exercise to verify that all new auxiliary fields are non-dynamical. The resulting (longer) action will be polynomial in the fields but their local symmetry transformations (generalizing those in \eqref{symmd}) will still contain the  factor $\frac 1{(\partial a)^2}$. So, starting from \cite{Pasti:1996vs} it was found more practical and simpler to work with the single auxiliary field $a(x)$ in the action.
}

\section{Conclusion}
In this paper we have shown that general results regarding the peculiar structure of  duality-invariant theories in four and six space-time dimensions do not extend to non-linear $10D$ chiral 4-form theories, because the number 81 of independent invariants in the latter is huge in comparison with single independent (4th-order) invariants in the former. This results in the existence of  a wide class of non-linear conformally invariant chiral 4-form theories in $D=10$. In contrast to the $4D$ and $6D$ non-linear duality-invariant theories which all can be viewed as stress-tensor ($T\overline T$-like) deformations of seed theories, in $D=10$, in general, the flow equations contain not only invariants of the stress-tensor, but also other invariant structures of $F^n$. The reason is that the number of independent invariants which one can construct from the components of $T_{\mu\nu}$ is 10, i.e. is much less than the number 81 of the independent invariants. It would be of interest to figure out whether there exist subclasses of non-linear $10D$ chiral 4-form theories which are generated exclusively by stress-tensor flow equations.\footnote{A similar problem in $D=8$ was addressed in \cite{Babaei-Aghbolagh:2020kjg}, where a possibility of obtaining a stress-tensor deformation of the free 3-form gauge field theory was demonstrated perturbatively up to the third order of the coupling parameter in the Lagrangian.} Another interesting issue is to understand how supersymmetry restricts the structure of 10th and higher order contributions of the self-dual five-form to the effective action of type IIB String Theory.

\section*{Acknowledgements}
The authors are grateful to Igor Bandos, Max Ba$\tilde {\rm n}$ados, Martin Cederwall and Chris Ferko for useful discussions and suggestions, and to Sergei Kuzenko for fruitful collaboration at early stages of this project and for useful comments on the manuscript.
The work of J.H. and D.S. is supported by the European Union under the Marie Sk\l{}odowska-Curie grant agreement number 101107602.\footnote{Views and opinions expressed are however those of the authors only and do not necessarily reflect those of the European Union or European Research Executive Agency. Neither the European Union nor the granting authority can be held responsible for them.}
D.S. was supported by the Swedish Research Council under grant no. 2021-06594 while participating in the research program  “Cohomological Aspects of Quantum Field Theory” at the  Institut Mittag-Leffler in Djursholm, Sweden on March 9-22, 2025, where part of this work has been done. The work of D.S. was also partially supported by CARIPARO Foundation under grant n. 68079,  by  the Australian Research Council Project DP230101629, the MCI, AEI, FEDER (UE) grant PID2021-125700NB-C21 “Gravity, Supergravity
and Superstrings” (GRASS) and  the Basque Government Grant IT1628-22.

\providecommand{\href}[2]{#2}\begingroup\raggedright\endgroup

\end{document}